\documentclass[twocolumn]{revtex4}
\usepackage[latin9]{inputenc}
\usepackage{amsmath}
\usepackage{graphicx}
\usepackage{wasysym}

\makeatletter
\@ifundefined{textcolor}{}
{%
 \definecolor{BLACK}{gray}{0}
 \definecolor{WHITE}{gray}{1}
 \definecolor{RED}{rgb}{1,0,0}
 \definecolor{GREEN}{rgb}{0,1,0}
 \definecolor{BLUE}{rgb}{0,0,1}
 \definecolor{CYAN}{cmyk}{1,0,0,0}
 \definecolor{MAGENTA}{cmyk}{0,1,0,0}
 \definecolor{YELLOW}{cmyk}{0,0,1,0}
}

\usepackage{float}\usepackage{amsfonts}

\makeatother

\begin{document}


\title{Breathing Mode of a Skyrmion on a Lattice}

\author{Dmitry A. Garanin$^{1}$, Reem Jaafar$^{2}$, and Eugene M. Chudnovsky$^{1}$}

\affiliation{$^{1}$ Physics Department, Herbert H. Lehman College and Graduate
School, The City University of New York, 250 Bedford Park Boulevard
West, Bronx, New York 10468-1589, USA\\
 $^{2}$Department of Mathematics, Engineering and Computer Science,
LaGuardia Community College, The City University of New York, 31-10
Thomson Avenue, Long Island City, NY 11101}

\date{\today}
\begin{abstract}
The breathing mode of a skyrmion, corresponding to coupled oscillations
of its size and chirality angle is studied numerically for a conservative
classical-spin system on a $500\times500$ lattice. The dependence
of the oscillation frequency on the magnetic field is computed. It
is linear at small fields, reaches maximum on increasing the field,
then sharply tends to zero as the field approaches the threshold above
which the skyrmion loses stability and collapses. Physically transparent
analytical model is developed that explains the results qualitatively
and provides the field dependence of the oscillation frequency that
is close to the one computed numerically. It is shown that a large-amplitude
breathing motion in which the skyrmion chirality angle $\gamma$ is
rotating in one direction is strongly damped and quickly ends by the
skyrmion collapse. To the contrary, smaller-amplitude breathing motion
in which $\gamma$ oscillates is undamped.
\end{abstract}
\maketitle

\section{Introduction}

\label{Intro}

Studies of skyrmions have opened a promising avenue for developing
new forms of memory storage and information processing \cite{Nagaosa2013,Tomasello,Zhang2015,Klaui2016,Leonov-NJP2016,Hoffmann-PhysRep2017,Fert-Nature2017}.
Skyrmions in thin films are defects of the uniformly magnetized ferromagnetic
state stabilized by topology. They had been first introduced in the
non-linear $\sigma$-model by Skyrme \cite{SkyrmePRC58} and later
intensively studied in nuclear physics \cite{Manton-book}. Their
topological properties in a two-dimensional (2D) Heisenberg exchange
model have been elucidated by Belavin and Polyakov (BP) \cite{BelPolJETP75,ec-book}.
In practice, topological stability of skyrmions that arises from the
continuous field model is violated in solids by the discreteness of
the atomic lattice \cite{cai12}. External magnetic field, magnetic
anisotropy, dipole-dipole interaction (DDI), thermal and quantum fluctuations,
etc., further break the symmetry of the exchange model, leading to
the uncontrolled collapse or expansion of skyrmions. For that reason
they are typically observed in non-centrosymmetric materials. In such
materials the Dzyaloshinskii-Moriya interaction (DMI) that arises
from the lack of the inversion symmetry provides stability of skyrmions
within a certain area of the phase diagram \cite{Leonov-NJP2016,buttner18}.
To date, stable isolated skyrmions have been experimentally observed
at room temperatures \cite{boulle16,moreau16,Fert-Nature2017}. It
has been demonstrated that the size of a skyrmion can be tuned by
the external magnetic field, with its radius shrinking on increasing
the field opposite to the skyrmion's spin until the skyrmion disappears
\cite{romming13,romming15}.

\begin{figure}
\centering{}\includegraphics[width=75mm]{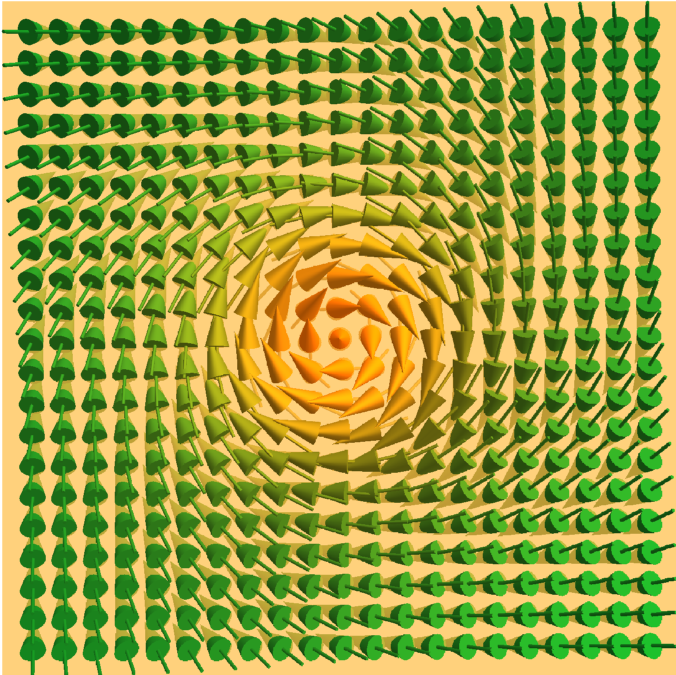} \caption{Spin field in a Bloch-type skyrmion studied in the paper. Its breathing
mode corresponds to the coupled oscillations of the skyrmion size
and spin angles.}
\label{Fig-skyrmion} 
\end{figure}

The shape of the smallest skyrmions is typically close to the shape
provided by the BP solution of the pure exchange model \cite{quantum},
while bigger skyrmions resemble magnetic bubbles studied in the past
\cite{Odell}. Field-theoretical approach to the internal dynamics
of skyrmions in nuclear physics and 2D magnets goes back to 1980s
\cite{Kaufluss1985,Ivanov1983}. More recently the interest to the
internal modes of skyrmions stabilized by the DMI has developed \cite{Garst2017}.
Chiral spin-wave modes of skyrmions and edge oscillations of skyrmion
bubbles, including a breathing-type mode, have been predicted \cite{Mochizuki-2012,Mahfudz2012}
and experimentally observed \cite{Onose2012} in skyrmion crystals.
A translational mode and different type of breathing modes have been
calculated for isolated skyrmions using Landau-Lifshitz-Gilbert (LLZ)
dynamics in Ref.\ \cite{Lin2014}. Hybridization of breathing modes
with quantized spin-wave modes in circular ultrathin magntic dots
have been investigated via micromagnetic computations \cite{Kim2014}.
Contribution of the internal modes to the mass of a skyrmion bubble
has been studied \cite{Lin2017,Kravchuk2018}. Most recently, LLZ
dissipative breathing dynamics of skyrmions and antiskyrmions has
been analyzed within Hamiltonian formalism \cite{Sinova2019}.

In this paper we focus on the problem that has not been previously
addressed: excitation spectrum of small skyrmions close to their stability
threshold. The breathing mode of a skyrmion shown in Fig.\ \ref{Fig-skyrmion}
is investigated. It corresponds to coupled oscillations of the skyrmion
size and spin angles. We will show that the frequency of the mode
has a distinct behavior in the collapse region that must be possible
to detect in experiment. It is linear on the field at weak fields,
riches maximum in the critical region and tends to zero as the field
approaches the collapse threshold. Close behavior has been obtained
by two independent methods. The first method consists of a purely
numerical computation of the spin dynamics on 2D lattices of size
ranging from $100\times100$ to $500\times500$, with a check that
the results are independent on the size in the limit of a large lattice.
The frequency of that internal skyrmion mode is below the spin-wave
spectrum of the uniformly magnetized ferromagnetic state, explaining
why no dissipation of the breathing mode has been observed in the
numerical experiment. The second method uses analytical and semi-analytical
models based upon Lagrangian dynamics of the skyrmion. It provides
a transparent physical picture of the behavior of the breathing mode
on the magnetic field. Quantitative agreement of the analytical method
with numerical calculation on the lattice is within $20\%$.

The paper is structured as follows. The model, the numerical method,
and the results for the small-amplitude skyrmion breathing mode computed
on a lattice are given in Section \ref{Sec-lattice}. Section \ref{Sec-analytical}
contains one crude qualitative approach and another more refined semi-analytical
approach to skyrmion internal oscillations that provide clear physical
interpretation of the numerical results. Dynamics of large-amplitude
breathing modes is considered in Sec. \ref{Sec-Large-amplitude-breathing-mode}.
Our conclusions are summarized in Section \ref{Sec-conclusions}.

\section{Skyrmion Breathing Mode in the Lattice Model}

\label{Sec-lattice} 

\subsection{General}

We consider a two-dimensional square lattice of normalized classical
spins, $\textbf{s}_{i}\equiv\textbf{S}_{i}/S$ where $\textbf{S}_{i}$
is a three-dimensional vector and $i=\{i_{x},i_{y}\}$ refers to the
lattice site. The Hamiltonian of the system is given by 
\begin{eqnarray}
\mathcal{H} & = & -\frac{S^{2}}{2}\sum_{ij}J_{ij}\textbf{s}_{i}\cdot\textbf{s}_{j}-HS\sum_{i}s_{iz}-\frac{DS^{2}}{2}\sum_{i}s_{iz}^{2}\nonumber \\
 & - & AS^{2}\sum_{i}\left[(\textbf{s}_{i}\times\textbf{s}_{i+\delta_{x}})\cdot\mathbf{e}_{x}+(\textbf{s}_{i}\times\textbf{s}_{i+\delta_{y}})\cdot\mathbf{e}_{y}\right].\label{energy-discrete}
\end{eqnarray}
The first term represents the Heisenberg exchange energy with the
exchange constant $J$ and sum is taken over the nearest neighbors.
The second term is the Zeeman interaction energy due to the external
field $H$ normal to the $xy$ plane. The third term is the energy
of the perpendicular magnetic anisotropy (PMA) of strength $D$. The
last term represents the Dzyaloshinskii-Moriya interaction (DMI) of
strength $A$, and $\textbf{s}_{i+\delta_{x}}=\textbf{s}_{i_{x}\pm1,i_{y}}$,
etc. For certainty, we have chosen the Bloch type DMI that favors
the Bloch-type skyrmions shown in Fig. \ref{Fig-skyrmion}.

The presence of the skyrmion in the system is revealed by a nonzero
topological charge: 
\begin{equation}
Q=\frac{1}{4\pi}\int dxdy\hspace{5pt}\textbf{s}\cdot\bigg(\frac{\partial\textbf{s}}{\partial x}\times\frac{\partial\textbf{s}}{\partial y}\bigg)\label{Q}
\end{equation}
that takes discrete values $Q=0,\pm1,\pm2,...$ In numerical work
we compute the discretized version of this expression.

Within the purely exchange continuous model, the BP solution for the
skyrmion with $Q=1$ and spins in the center of the skyrmion pointing
up against the spin-down background in terms of polar coordinates
$x=r\cos\phi$, $y=r\sin\phi$ has the form
\begin{equation}
\left\{ \begin{array}{c}
s_{x}\\
s_{y}
\end{array}\right\} =\frac{2\lambda r}{r^{2}+\lambda^{2}}\left\{ \begin{array}{c}
\cos(\phi+\gamma)\\
\sin(\phi+\gamma)
\end{array}\right\} ,\quad s_{z}=\frac{\lambda^{2}-r^{2}}{\lambda^{2}+r^{2}}.\label{skyrmion-components}
\end{equation}
Here $\lambda$ is the skyrmion size and spins are rotated away from
the radial direction by the chirality angle $\gamma$. The energy
of the skyrmion is independent of $\lambda$ and $\gamma$ and equal
to $4\pi JS^{2}$ above that of the uniform state. This is the invariance
found by Belavin and Polyakov.

It was shown that the discreteness of the lattice makes the energy
decrease with decreasing $\lambda$ that leads to the skyrmion collapse
\cite{cai12}. Other interactions apart from the exchange, also break
the invariance. The PMA leads to the energy increase with $\lambda$
thus it should lead to a collapse. However, the dipole-dipole interaction
favors the skyrmion expansion and, together with the PMA, it can stabilize
the skyrmion at a particular size. DMI favors skyrmion expansion and
adjustment of the chirality angle to a particulr value ($\gamma=\pi/2$
for the Bloch DMI with $A>0$). This expansion can be limited by the
magnetic field applied in the negative direction with respect to the
skyrmion's spin. This stabilizes the skyrmion at a particular size.
If the applied field becomes too strong, the skyrmion collapses.

Certainly, the shape of the skyrmion stabilized by non-exchange interactions
differs from the BP shape. However, since the exchange is the strongest
interaction, at least small skyrmions are only weakly distorted. Thus,
if makes sense to use Eq. (\ref{skyrmion-components}) as the Ansatz
in the analytical approach. Below, we will ignore the DDI and mainly
investigate the model with the DMI, numerically and analytically,
focusing on the breathing mode.

Breathing mode is the lowest-frequency local mode of the skyrmion
in which the skyrmion size is oscillating around its equilibrium value.
As we will see, this is accompanied by oscillations of the dynamically
conjugate variable, the chirality angle $\gamma$. There should be
faster modes including various deformations of the skyrmion, that
will not be investigated. 

\subsection{Numerical energy minimization and the skyrmion size}

To find the frequency of the breathing mode numerically, first the
energy minimization was done for a particular set of parameters. The
main choice was $A/J=0.02$, whereas the applied field $H$ changed
between its collapse value and zero. As the initial condition, any
bubble with $Q=1$ at the center of the system can be used. The numerical
method \cite{DCP-PRB2013} combines sequential rotations of spins
${\bf s}_{i}$ towards the direction of the local effective field,
${\bf H}_{{\rm eff},i}=-\partial{\cal H}/\partial{\bf s}_{i}$, with
the probability $\alpha$, and the energy-conserving spin flips (overrelaxation),
${\bf s}_{i}\to2({\bf s}_{i}\cdot{\bf H}_{{\rm eff},i}){\bf H}_{{\rm eff},i}/H_{{\rm eff},i}^{2}-{\bf s}_{i}$,
with the probability $1-\alpha$. We used $\alpha=0.03$ that ensures
the fastest relaxation. It was found that the breathing mode can be
seen only in the model with periodic boundary conditions (pbc), both
for the exchange and for the DMI. In the case of free boundary conditions,
there are surface modes that interfere in the extraction of the frequency
of the breathing mode. Thus, all computation were performed on the
model with pbc. 

\begin{figure}
\begin{centering}
\includegraphics[width=8.8cm]{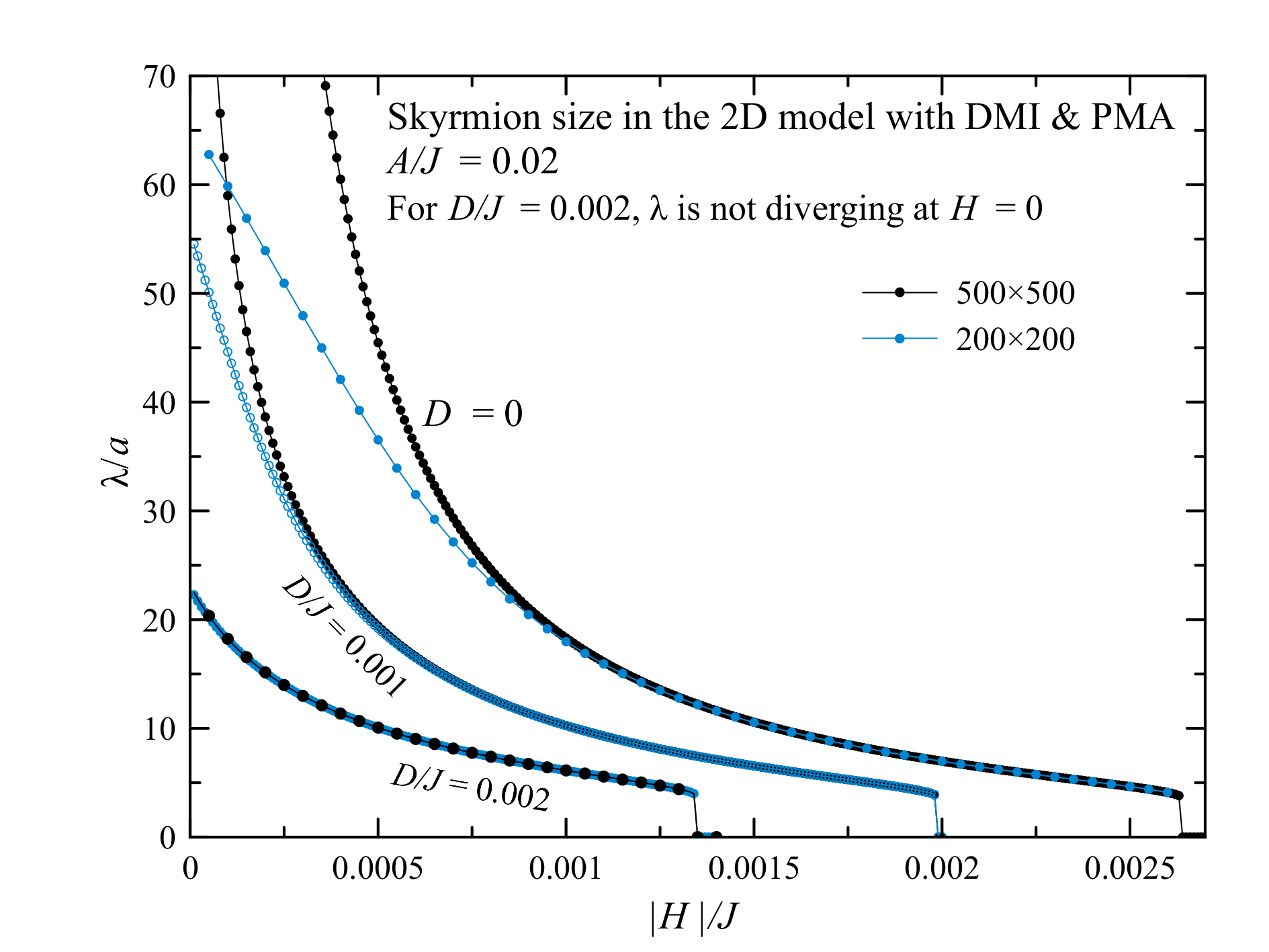}
\par\end{centering}
\caption{Skyrmion size vs the applied magnetic field for $A/J=0.02$ and different
PMA values.}

\label{Fig-lambda_vs_absH}
\end{figure}

The skyrmion size $\lambda$ can be extracted from the numerical data
as \cite{cai12} 
\begin{equation}
\lambda_{n}^{2}=\frac{n-1}{2^{n}\pi}a^{2}\sum_{i}\left(s_{iz}+1\right)^{n},
\end{equation}
in our case $s_{iz}=-1$ in the background and $s_{iz}=1$ at the
center of the skyrmion. For the BP skyrmions with $s_{z}$ given by
Eq. (\ref{skyrmion-components}), one has $\lambda_{n}=\lambda$ for
any $n$. In this paper, we used $\lambda=\lambda_{4}$ to represent
the numerically computed skyrmion size. We also computed the components
of the average spin of the system as
\begin{equation}
\mathbf{m}=\frac{1}{N}\sum_{i}\mathbf{s}_{i}.\label{m_Def}
\end{equation}
One can also define the skyrmion spin as 
\begin{equation}
\mathcal{M}=\sum_{i}\left(s_{iz}+1\right).\label{M_Def}
\end{equation}
The angle $\gamma$ was extracted by building the sum of dot products
of the lattice spins $\mathbf{s}_{i}$ by the radial vectors $\mathbf{r}_{i}$
(with respect to the center of the lattice), that yields $\cos\gamma$,
and by the $\boldsymbol{\phi}_{i}$-vectors, that are perpendicular
to the $\mathbf{r}_{i}$-vectors and point counterclockwise, to find
$\sin\gamma$. 

The results for the equilibrium skyrmion size vs $H$ for two different
system sizes, $A/J=0.02$, and three different values of the PMA,
$D/J=0$, 0.001, and 0.002, are shown in Fig. \ref{Fig-lambda_vs_absH}.
For $D=0$, the skyrmion size diverges in the limit $H\rightarrow0$.
However, the divergence is limited by the system size that is clearly
seen in the figure. Thus, for a small field, a large system size is
needed. PMA tends to decrease the skyrmion size, thus for $D/J=0.001$
the latter is noticeably smaller, while the collapse field is smaller,
too. The skyrmion size still diverges for $H\rightarrow0$. To the
contrary, for a stronger PMA, $D/J=0.002$, the skyrmion size does
not diverge and the results for both system sizes are the same. This
means that a sufficiently strong PMA can stabilize the skyrmion at
$H=0$. On the other hand, if the PMA becomes too strong, the skyrmion
will collapse. Thus, there is a range of $D$ that stabilize the skyrmion
at $H=0$.

\begin{figure}
\centering{}\includegraphics[width=88mm]{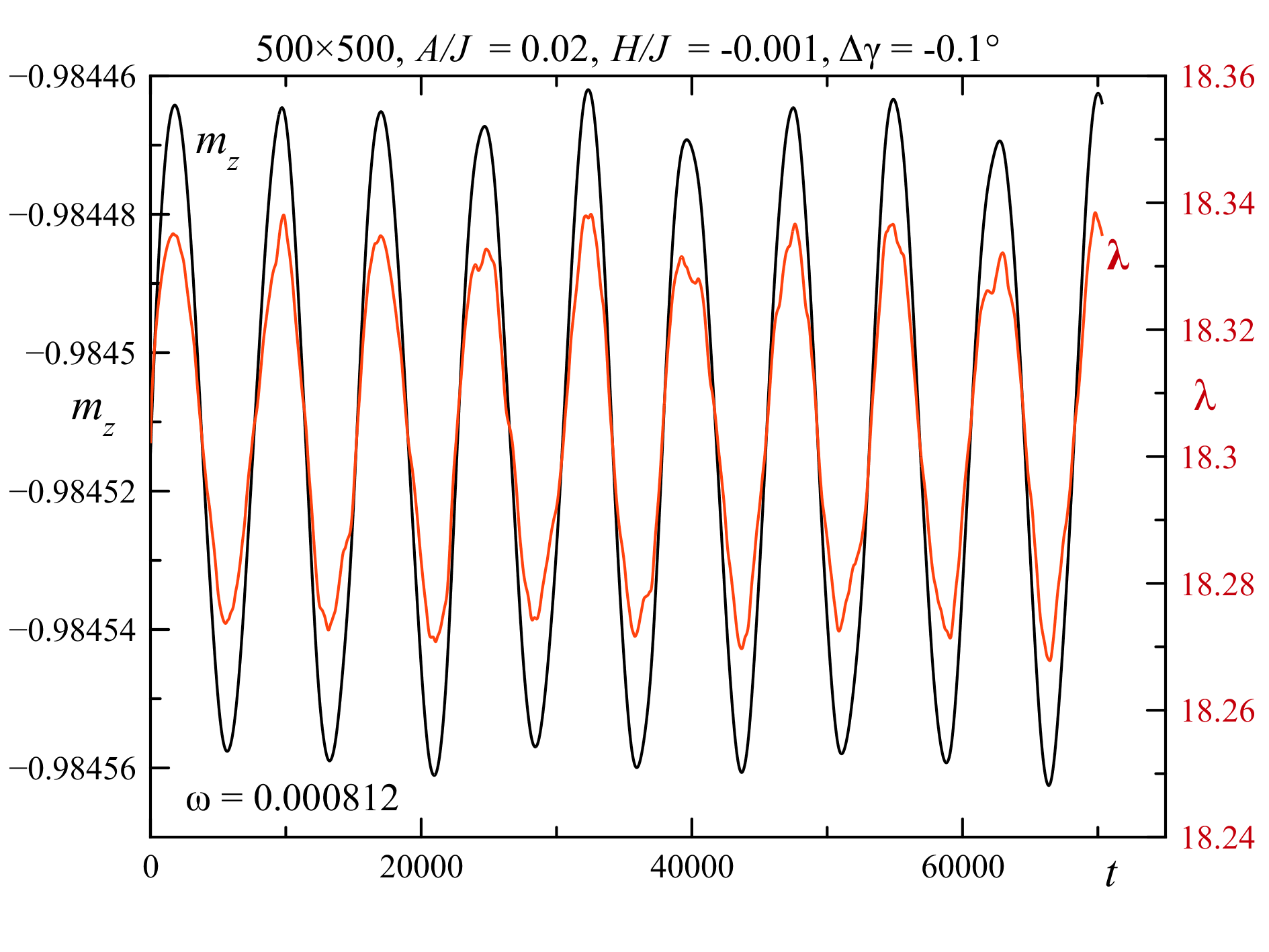} \caption{Numerically obtained oscillations of $m_{z}$ and of the skyrmion
size $\lambda$ in the breathing mode. }
\label{Fig-mz_and_lambda_oscillations} 
\end{figure}

\subsection{Numerical dynamics of the breathing mode}

After the equilibrium skyrmion configuration was found, the frequency
of its oscillations around the equilibrium was measured by running
the dynamical evolution following the rotation of all spins in the
system by $\Delta\gamma=1\textdegree$ around the $z$-axis. We used
the fourth-order Runge-Kutta ordinary-differential-equation solver
with the integration step $0.2$ in the units of $\hbar/J$ to solve
the system of Larmor equations of motion $\hbar\dot{\mathbf{s}}_{i}=\mathbf{s}_{i}\times{\bf H}_{{\rm eff},i}$
for the lattice spins. No damping was included in this computation.
For the small DMI constant and the applied field used here, the dynamics
is rather slow, so that the discretization error of the Runge-Kutta
method is rather small. However, one cannot significantly increase
the step as already for the step $0.3$ an instability occurs due
to the exchange term in the equations.

\begin{figure}
\begin{centering}
\includegraphics[width=88mm]{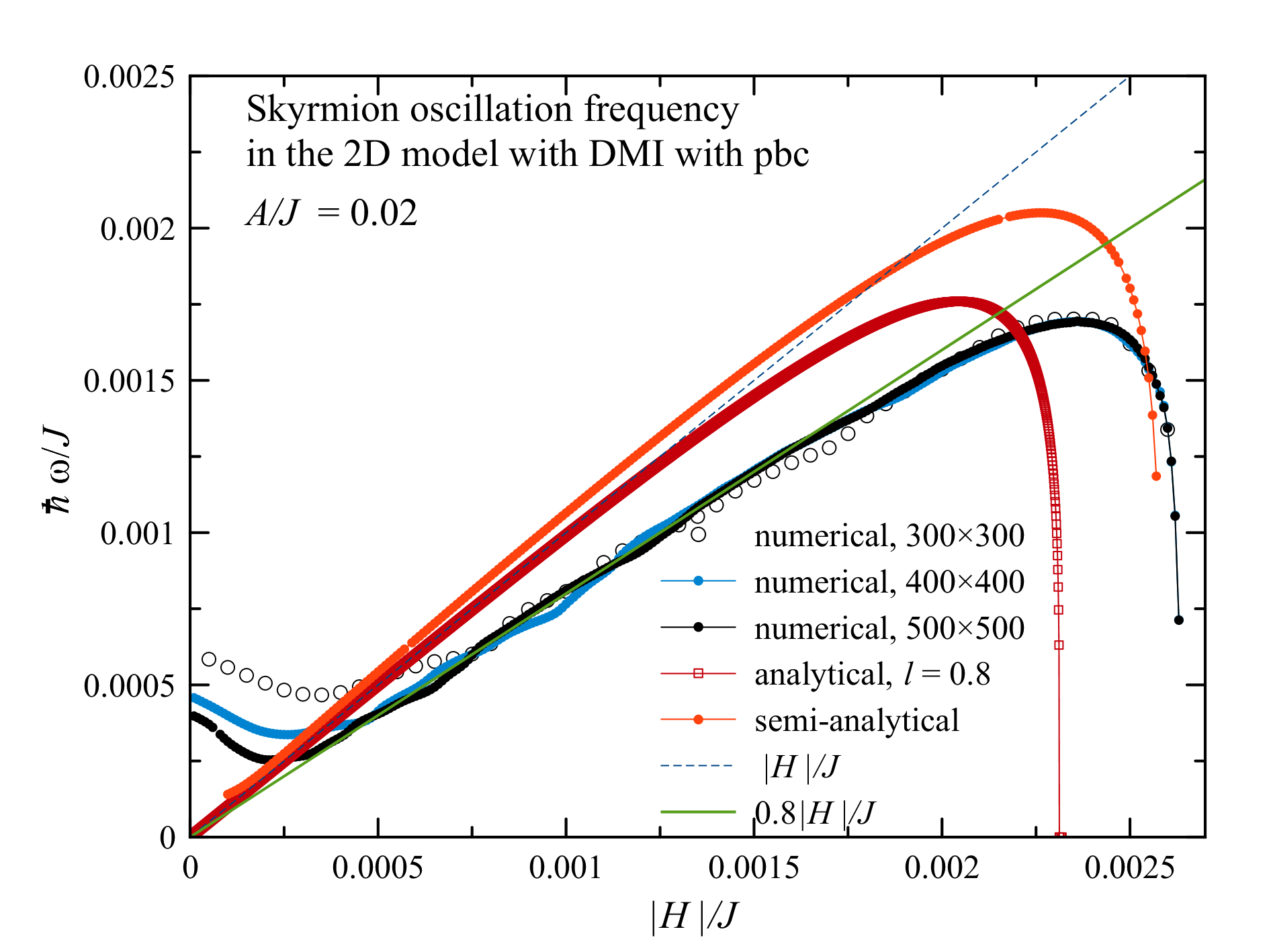} 
\par\end{centering}
\centering{}\caption{Field dependence of the oscillation frequency of the breathing mode
computed numerically on lattices of different sizes, analytically
for a BP skyrmion shape with logarithmic accuracy, and semi-analytically
in a model with the skyrmion shape corrected, see Section \ref{Sec-analytical}.}
\label{Fig-omega-H+theory} 
\end{figure}

The computation was done independently for each value of $H$ in parallel
using Worfram Mathematica with vectorization and compilation on a
20-core Dell Precision Workstation (16 cores used by Mathematica).
Computations performed for $A/J=0.1$, $0.02$, $0.01$ show qualitatively
similar behavior. In the paper the results are given for $A/J=0.02$
with $D/J=0$ and $D/J=0.002$. They were computed for the system
sizes $100\times100$, $200\times200$, $300\times300$, $400\times400$,
and $500\times500$. A greater system size is needed for small applied
fields when the skyrmion size becomes large. Comparison of the results
for different system sizes show that at our maximal size $500\times500$
there are no finite-size effects in the main range of $H$, except
for the smallest $H$.

It was found that the skyrmion size $\lambda$ and the system's average
spin $m_{z}$ (or, equivalently, the skyrmion's magnetic moment $\mathcal{M}$)
performed periodic oscillations with a weak anharmonicity, see Fig.\ \ref{Fig-mz_and_lambda_oscillations}.
Since the curves for $m_{z}$ are smoother than those for $\lambda$,
the former were used to extract the oscillation frequency. The anharmonicity
could be attributed to a weak hybridization of the breathing mode
with the other local modes (see, e.g., Ref. \cite{dessuekimsta18prb})
that also could be excited by rotating all spins by $\Delta\gamma$.
Indeed, the deviation from the equilibrium skyrmion state can be expanded
over the set of local modes. In this expansion, the breathing mode
should be the strongest while other modes enter with smaller weights
and thus they distort the breathing dynamics to some extent seen in
the dependences $\lambda(t)$ and $m_{z}(t)$. On the other hand,
no damping of the breathing mode was detected. This can be explained
by the fact that its frequency always falls below the frequency of
the uniform precession, that is, the breathing mode is not resonating
with the spin-wave band. The time interval between the adjacent maximum
and minimum (or between a minimum and a maximum) of $m_{z}$ was interpreted
as the half of the period to extract the oscillation frequency. This
half-period of the evolution costs much more computer time than finding
the skyrmion's equilibrium state in the first stage. In the parallelized
computation for diffeent values of $H$, after reaching the first
maximum and the first minimum of $m_{z}$, the computation was terminated
and the frequency was recorded.

\begin{figure}
\begin{centering}
\includegraphics[width=8.8cm]{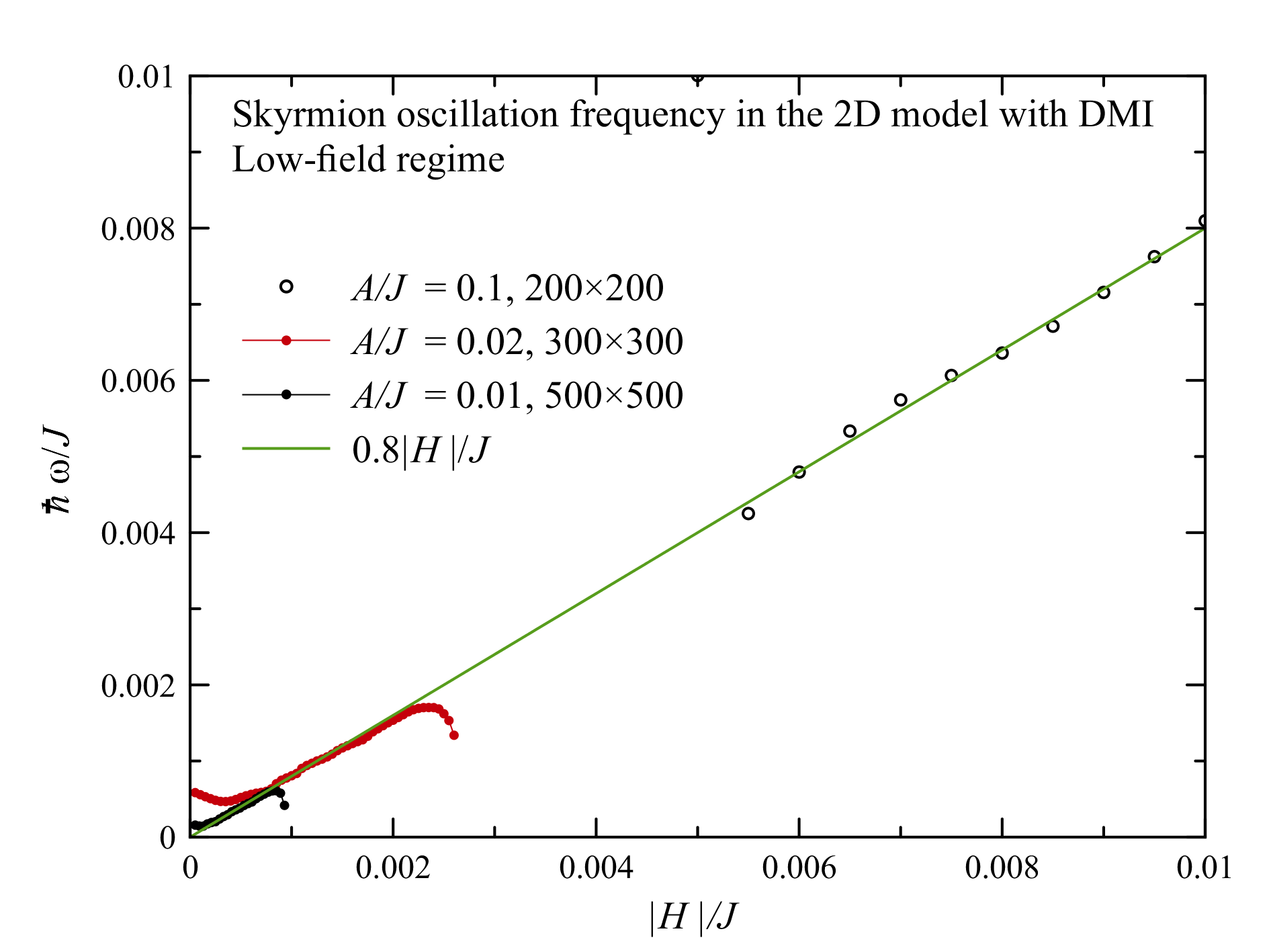} 
\par\end{centering}
\caption{Frequency of the breathing mode in the linear-$H$ regime for different
values of the DMI constant $A$.}

\label{Fig-omega_vs_H_DDM_Bloch_pbc_weak_H}
\end{figure}

\begin{figure}
\begin{centering}
\includegraphics[width=8.8cm]{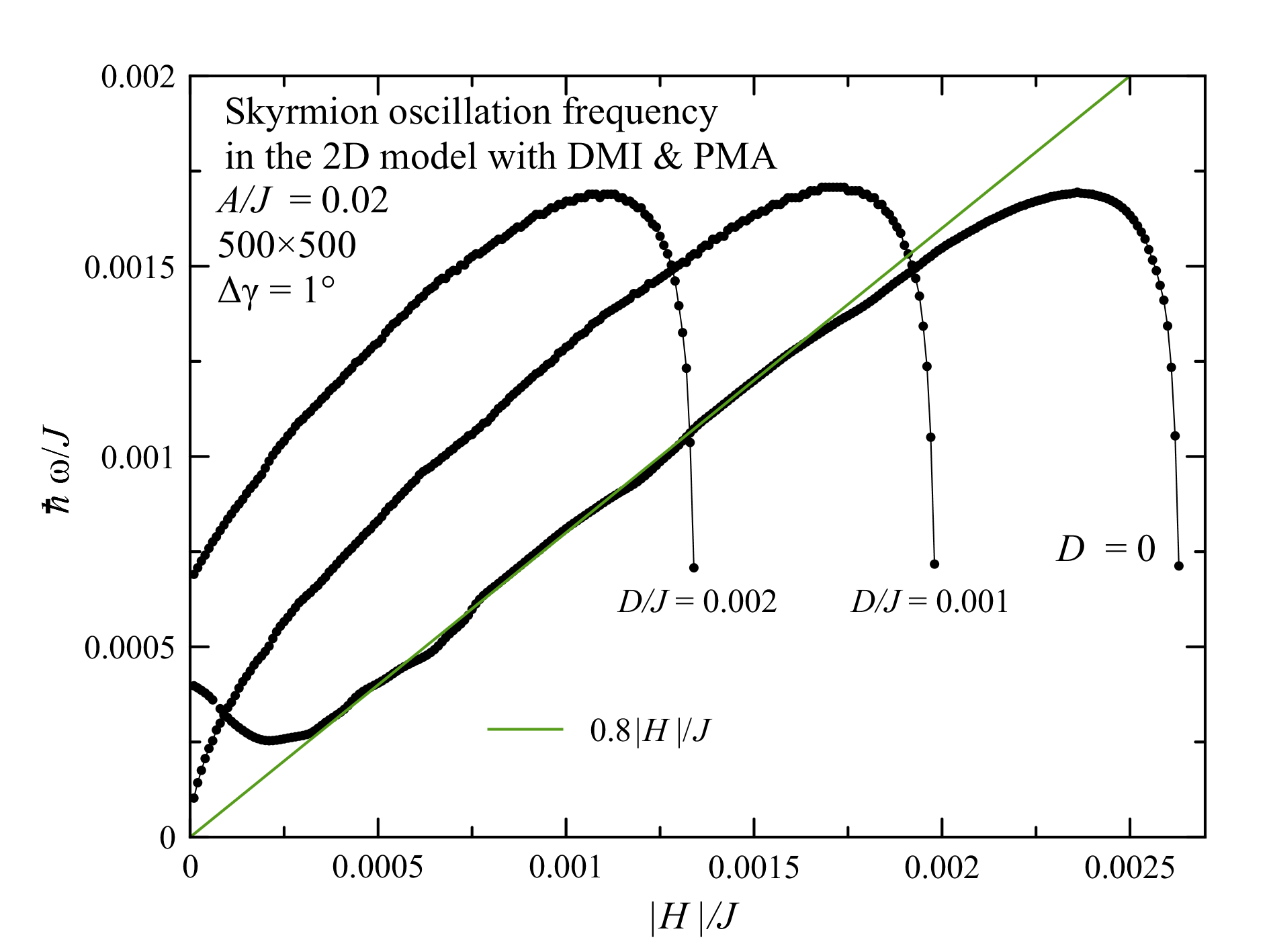}
\par\end{centering}
\caption{Breathing-mode frequency vs the applied magnetic field for $A/J=0.02$
and different PMA values for the $500\times500$ system size.}

\label{Fig-omega_vs_absH_PMA}
\end{figure}

The dependence of the frequency of the breathing mode on the magnetic
field at $D=0$ is shown in Fig.\ \ref{Fig-omega-H+theory}. The
$\omega(H)$ curves obtained numerically on lattices of different
size have little size dependence, except for weak fields, where $\omega(H)$
has a size-dependent uptick. Here one can expect skyrmions branching
out and transform to a laminar domain state. The $\omega(H)$ curves
exhibit a characteristic maximum on approach to the critical field
above which the skyrmion collapses. Here $\omega(H)$ goes to zero
steeply. On the left side of the maximum, where skyrmions are big
and the lattice discreteness becomes unimportant, $\omega(H)$ goes
apparently linearly and can be approximated by the dependence $\hbar\omega(H)=0.8H$.
Qualitatively similar behavior is exhibited by the $\omega(H)$ curves
computed analytically in the next Section, that are shown in the same
figure. 

It is remarkable that the dependence $\hbar\omega(H)=0.8H$ holds
for different values of the DMI constant $A$. Thus, the coefficient
0.8 is a universal number.

Numerical results in the presence of the PMA are shown in Fig.\ \ref{Fig-omega_vs_absH_PMA}
for $A/J=0.02$ and the system size $500\times500$. In accordance
with Fig. \ref{Fig-lambda_vs_absH}, the collapse fields and the entire
curves shift to the left with increasing the PMA. The breathing-mode
frequency $\omega(H)$ is well below the FMR frequency $\omega_{\mathrm{FMR}}=\left(D+H\right)/\hbar$.
For $D/J=0.001$, $\omega(H)$ goes to zero with a high slope at $H\rightarrow0$.
This must be related to the divergence of $\lambda(H)$ seen in Fig.
\ref{Fig-lambda_vs_absH}. To the contrary, for $D/J=0.002$, the
breathing-mode frequency remains finite at $H=0$ that correlates
with the finite $\lambda$ at zero field in Fig. \ref{Fig-lambda_vs_absH}.

\begin{figure}
\begin{centering}
\includegraphics[width=8.8cm]{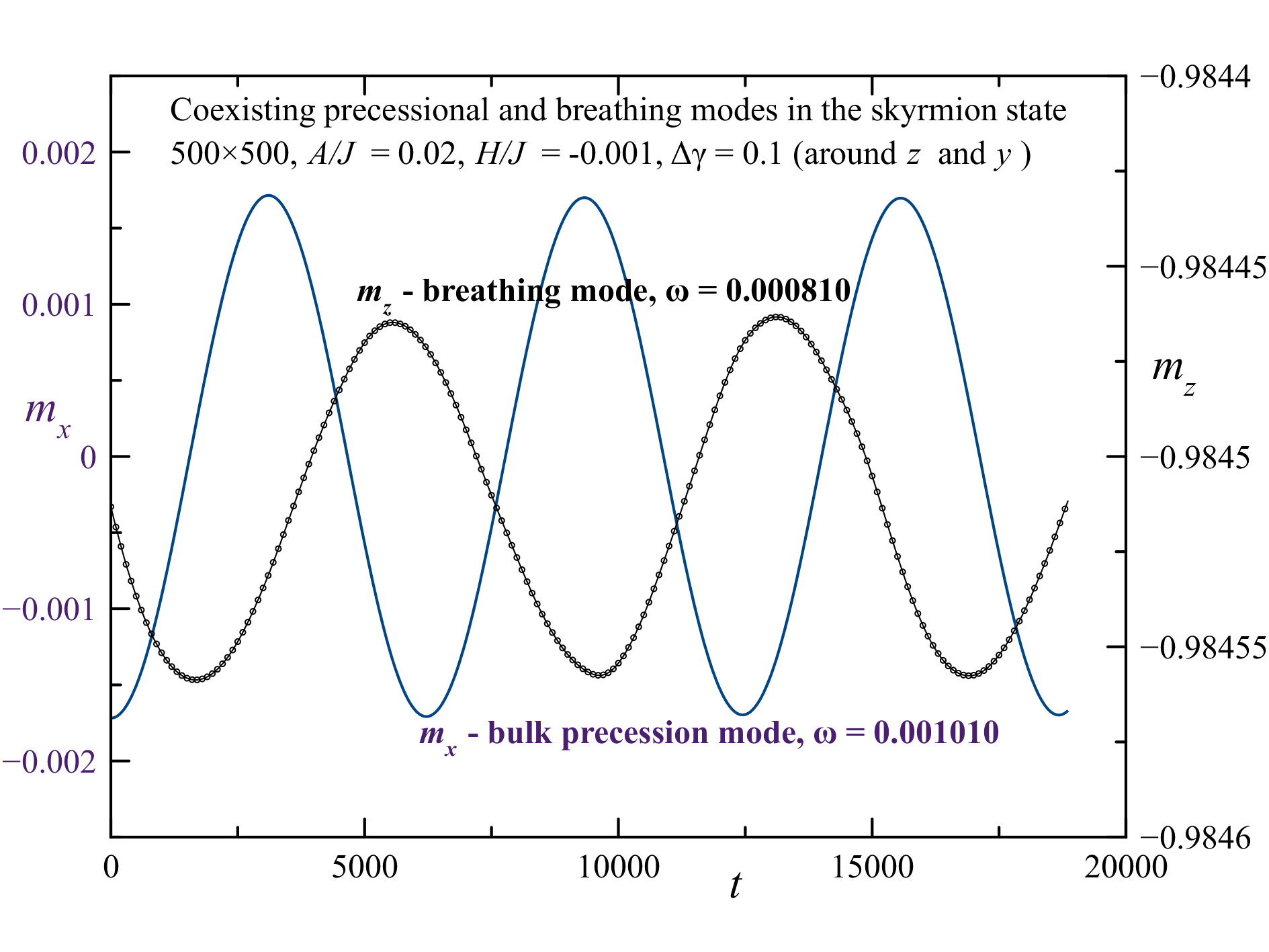}
\par\end{centering}
\caption{Coexisting breathing and bulk precession modes after rotating spins
around $z$- and $y$-axes out of the equilibrium-skyrmion state.
These modes oscillate at close but distinctly different frequences.}

\label{Fig-Coexisting_modes}
\end{figure}

It is non-trivial that at $D=0$ the frequency of the breathing mode
in the main region of the applied field follows, for any value of
$A$, a linear law $\hbar\omega(H)=0.8H$ that is resembling that
for the FMR frequency $\hbar\omega_{\mathrm{FMR}}=H$ but has a smaller
coefficient. In fact, the breathing mode and the bulk-precession mode
have totally different structures. In the precession mode, $m_{z}=\mathrm{const}$
while $m_{x}$ and $m_{y}$ are precessing. In the breathing mode,
$m_{z}$ is oscillating, while the spin orientation quantified by
the angle $\gamma$ is performing small oscillations around its equilibrium
value. These two main modes are independent at small amplitudes and
can coexist. To check this, we initiated dynamics by rotating all
spins by $0.1\textdegree$ around the $y$- and $z$-axes. This excites
both modes, the temporal evolution of which is shown in Fig. \ref{Fig-Coexisting_modes}.
One can see that the frequencies of these modes are different. Moreover,
rotation of spins only around the $z$-axis, to excite the breathing
mode, also excites the bulk precession mode with a very small, although
numerically detectable amplitude, evolving with the proper FMR frequency.

\section{Skyrmion Breathing Mode in a Spin-Field Model}

\label{Sec-analytical}

\subsection{Analytical approach with the BP Ansatz for the skyrmion shape}

\begin{figure}
\centering{}\includegraphics[width=80mm]{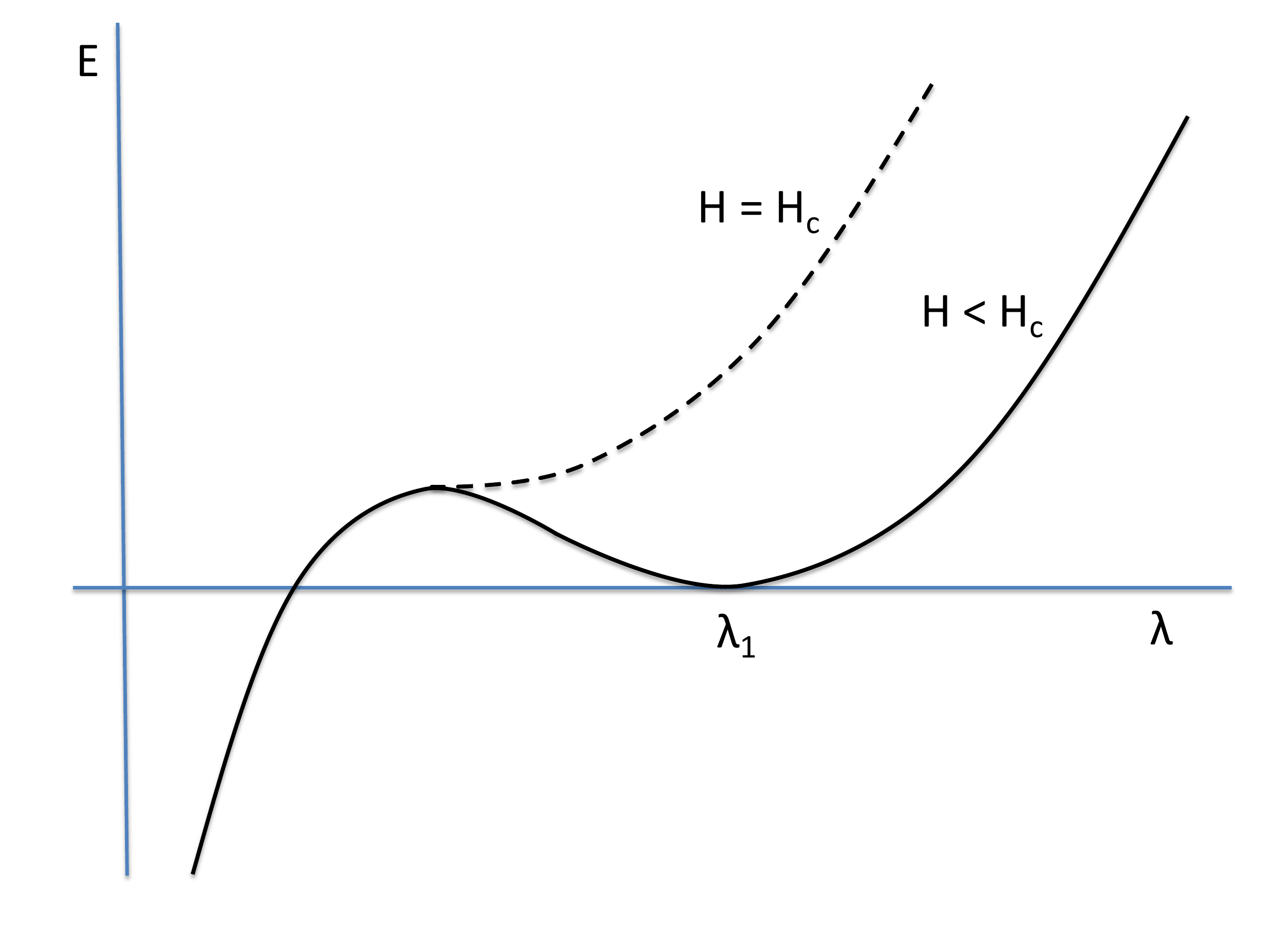}\caption{Dependence of the skyrmion energy on skyrmion size $\lambda$ for
$\gamma=\pi/2$. The minimum corresponding to the equilibrium size
disappears for $H>H_{c}$. At $H=H_{c}$ the energy has an inflection
point where both first and second derivative are zero.}
\label{Fig-energy} 
\end{figure}

The fact that $\omega(H)$ tends to zero at the critical field, $H=H_{c}$,
can be easily understood from the dependence of the energy on the
field and the size of the skyrmion shown in Fig.\ \ref{Fig-energy}.
Skyrmion of a stable size exists at $H>H_{c}$ when $E(\lambda)$
has a minimum. At $H=H_{c}$ the minimum becomes an inflection point
where both, first and second derivative of $E$ on $\lambda$, are
zero. It is the nullification of the second derivative that makes
the frequency of the breathing mode zero at $H=H_{c}$. In this Section
we develop analytical approach that elucidates the dynamics of the
breathing mode in simple physical terms.

The continuous analog of Eq.\ (\ref{energy-discrete}) is 
\begin{eqnarray}
{\cal H} & = & \frac{JS^{2}}{2}\int dxdy\left[\left(\frac{\partial{\bf s}}{\partial x}\right)^{2}+\left(\frac{\partial{\bf s}}{\partial y}\right)^{2}\right]\nonumber \\
 & - & \frac{JS^{2}a^{2}}{24}\int dxdy\left[\left(\frac{\partial^{2}{\bf s}}{\partial x^{2}}\right)^{2}+\left(\frac{\partial^{2}{\bf s}}{\partial y^{2}}\right)^{2}\right]\nonumber \\
 & - & \frac{HS}{a^{2}}\int dxdy\,s_{z}-\frac{DS^{2}}{2a^{2}}\int dxdy\,s_{z}^{2}\nonumber \\
 & + & \frac{AS^{2}}{a}\int dxdy\left[\left({\bf s}\times\frac{\partial{\bf s}}{\partial x}\right)\cdot\mathbf{e}_{x}+\left({\bf s}\times\frac{\partial{\bf s}}{\partial y}\right)\cdot\mathbf{e}_{y}\right].\label{E-continuous}
\end{eqnarray}
The second term in this expression arises from taking into consideration
the next derivatives in the expansion of the discrete form of the
exchange energy that dominates spin interactions at small distances.
The spin combination in the DMI energy can be rewritten as $\mathbf{s}\cdot\left(\nabla\times\mathbf{s}\right)$.
For a small skyrmion that is close to the BP shape, substitution of
Eq.\ (\ref{skyrmion-components}) into Eq.\ (\ref{E-continuous})
gives at $D=0$ 
\begin{equation}
\bar{E}\equiv\frac{{\cal H}}{4\pi JS^{2}}=h\bar{\lambda}^{2}l(\bar{\lambda})-\frac{1}{6\bar{\lambda}^{2}}-\alpha\bar{\lambda}\sin\gamma,\label{E-l}
\end{equation}
where $h\equiv H/(JS)$, $\alpha\equiv A/J$, $\bar{\lambda}\equiv\lambda/a$,
and $l(\bar{\lambda})$ has a logarithmic dependence on $\lambda$
that is sensitive to the shape of the skyrmion far from its center.
This function for $\gamma=\pi/2$ is shown qualitatively in Fig. \ref{Fig-energy}. 

\begin{figure}
\centering{}\includegraphics[width=80mm]{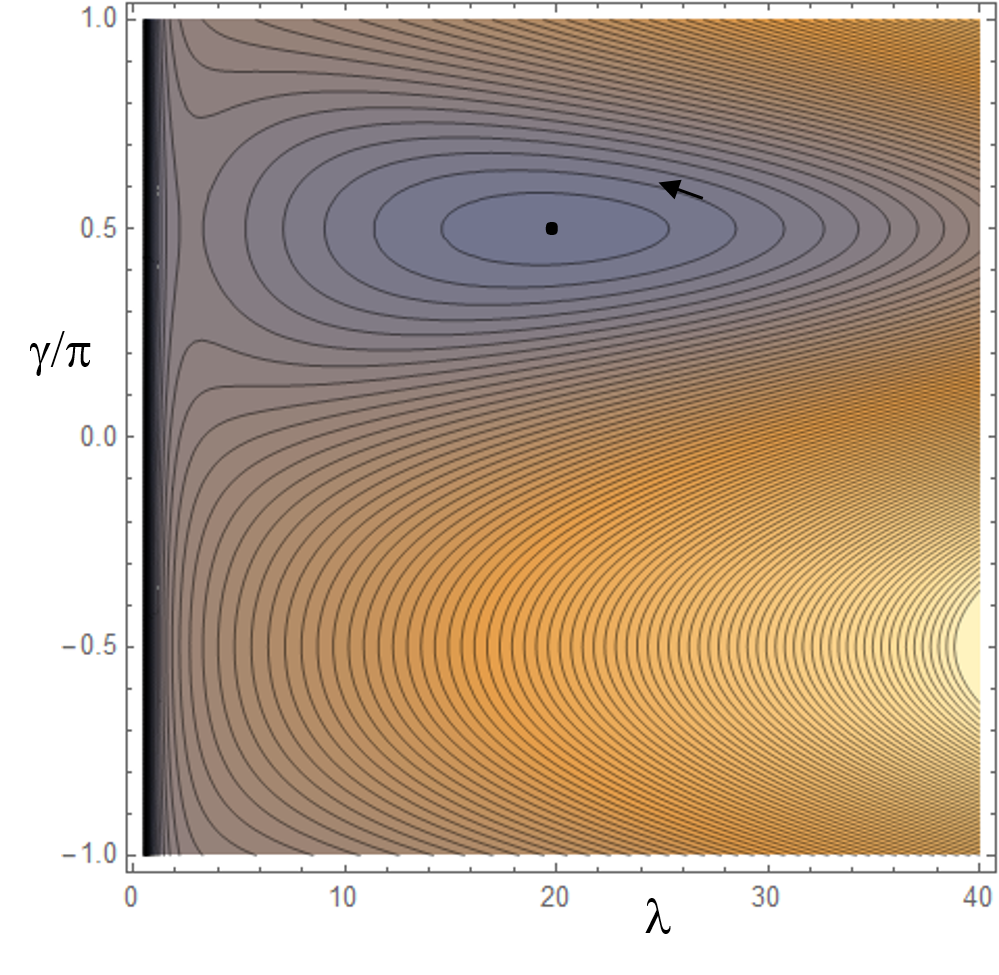}\caption{Equipotential lines in the $\left\{ \lambda,\gamma\right\} $ plane
corresponding to Eq.\ (\ref{E-l}) with $l={\rm const}$.}
\label{Fig-equipotential} 
\end{figure}

Consider first a crude approximation with $l={\rm const}$. The whole
potential landscape of Eq. \ref{E-l} is shown in Fig. \ref{Fig-equipotential}
for $\alpha=0.02$, $h=-0.001$, and $l=0.5$. As the dynamics of
the skyrmion conserves the energy, the system is moving in its phase
space $\left\{ \lambda,\gamma\right\} $ along the equipotential lines.
The motion occurs in the counterclockwise direction. The center point
is the energy minimum $\gamma=\pi/2$ and $\lambda=\lambda_{1}$ defined
by the angebraic equation
\begin{equation}
\frac{\partial\bar{E}}{\partial\bar{\lambda}}=2lh\bar{\lambda}+\frac{1}{3\bar{\lambda}^{3}}-\alpha=0.\label{min}
\end{equation}
There are two regimes of the breathing motion of the skyrmion, if
one looks at the behavior of $\gamma$. The oscillating regime corresponds
to closed trajectories around the metastable energy minimum. The rotating
regime is described by $\gamma$ steadily increasing with time. However,
one can reduce $\gamma$ to the interval $\left(-\pi,\pi\right)$.
In this representation, in the rotating regime trajectories are leaving
the area through the top and reentering through the bottom. The skyrmion
size $\lambda$ is oscillating in both regimes.

Solution of Eq.\ (\ref{min}) together with the equation $\partial^{2}\bar{E}/\partial\bar{\lambda}^{2}=2lh-1/\bar{\lambda}^{4}=0$
gives the critical field, $h_{c}$, and the value of $\bar{\lambda}_{1}=\bar{\lambda}_{c}=$
at the critical field, 
\begin{equation}
h_{c}=\frac{1}{2l}\left(\frac{3\alpha}{4}\right)^{4/3},\qquad\bar{\lambda}_{c}=\left(\frac{4}{3\alpha}\right)^{1/3}.\label{crit}
\end{equation}
At $h\ll h_{c}$ one has $\bar{\lambda}_{1}\approx\alpha/(2lh)$.
This roughly agrees with the lattice results shown in Fig.\ \ref{Fig-omega-H+theory}.

To study the dynamics of the spin system, consider the Lagrangian
\cite{quantum} 
\begin{equation}
{\cal L}=\hbar S\int dxdy\dot{\Phi}(\cos\Theta+1)-{\cal H},\label{Lagrangian}
\end{equation}
where $\Theta$ and $\Phi$ are spherical coordinates of ${\bf S}$,
satisfying $\cos\Theta+1=s_{z}+1$ and $\tan\Phi=s_{y}/s_{x}$. Substituting
here Eq.\ (\ref{skyrmion-components}), which results in $\dot{\Phi}\equiv d\Phi/dt=\dot{\gamma}$,
upon integration one obtains 
\begin{equation}
{\cal L}=4\pi\hbar S\dot{\gamma}\bar{\lambda}^{2}l-E(\bar{\lambda},\gamma).
\end{equation}
The Euler-Lagrange equations are 
\begin{equation}
\frac{\partial{\cal L}}{\partial\bar{\lambda}}=0,\qquad\frac{d}{dt}\frac{\partial{\cal L}}{\partial\dot{\gamma}}=\frac{\partial{\cal L}}{\partial\gamma},\label{EL}
\end{equation}
resulting in the coupled equations of motion for $\bar{\lambda}$
and $\gamma$: 
\begin{eqnarray}
2\frac{d\gamma}{dt}\bar{\lambda}l & = & -\alpha\sin\gamma+2h\bar{\lambda}l+\frac{1}{3\bar{\lambda}^{3}}\label{motion}\\
2\frac{d\lambda}{dt}l & = & \alpha\cos\gamma.
\end{eqnarray}
Linearization of the above equations for small amplitude oscillations
yields the frequency of the breathing mode, 
\begin{equation}
\bar{\omega}_{1}(h)=\frac{\hbar\omega}{JS}=\frac{\alpha^{1/2}}{2l\bar{\lambda}_{1}^{1/2}}\sqrt{2lh-\frac{1}{\bar{\lambda}_{1}^{4}}},\label{omega1lambda1}
\end{equation}
where $\bar{\lambda}_{1}(h)$ is given by Eq.\ (\ref{min}). Its
dependence on the magnetic field for $l=0.8$ is shown in Fig.\ (\ref{Fig-omega-H+theory}).

It is easy to see that at $h\ll h_{c}$ the above equation gives $\bar{\omega}_{1}=h[1-(\bar{\lambda}_{c}/\bar{\lambda}_{1})^{3}/4]\approx h$.
This coincides with the FMR frequency and differs from a more accurate
numerical lattice result $\bar{\omega}_{1}\approx0.8h$ that brings
the frequency of the breathing mode below the bottom of the spin-wave
spectrum in the bulk. The latter does not allow the breathing mode
to decay into spin waves and is responsible for its non-dissipative
dynamics if damping from other sources is not introduced by hand into
the equations of motion.

The above method relies on a fitting parameter $l$ to come close
to the numerically obtained critical field. Although this simple method
provides a physical picture of the breathing-mode dynamics and provides
qualitatively correct results including the maximim of the breathing-mode
frequency, there are discrepancies with the numerical solution both
near the skyrmion collapse and at low fields.

\subsection{Semi-analytical approach using corrected skyrmion shape}

A better approximation not using any fitting parameters can be developed
if one takes into account the deformation of the BP shape of the skyrmion
at large distances. Because of the applied field $H$, the spin field
approaches its background value $-1$ exponentially at the magnetic
length $\delta_{H}=\sqrt{JS/|H|}$. In the limit $\lambda\ll\delta_{H}$,
the asymptotic solution of the linearized equation for $\mathbf{s}(\mathbf{r})$
at $r\gg\lambda$ can be combined with the BP solution at $r\ll\delta_{H}$.
This leads to replacement of Eq.\ (\ref{skyrmion-components}) by
\begin{equation}
\left\{ \begin{array}{c}
s_{x}\\
s_{y}
\end{array}\right\} =\frac{2\lambda f(r)}{f^{2}(r)+\lambda^{2}}\left\{ \begin{array}{c}
\cos(\phi+\gamma)\\
\sin(\phi+\gamma)
\end{array}\right\} ,\quad s_{z}=\frac{\lambda^{2}-f^{2}(r)}{\lambda^{2}+f^{2}(r)},\label{skyrmion-components-modified}
\end{equation}
where $f(r)=\delta_{H}/K_{1}(r/\delta_{H})$ and $K_{1}$ is the MacDonald
function (see, e.g, Ref. \cite{Ivanov1983}). Although formally valid
for $\lambda\ll\delta_{H}$, this solution that rescales the distance
from the skyrmion's center proves to be remarkably robust and provides
good results in a wide range of $H$. The reason is that the actual
skyrmion profile obtained numerically or within this approximation,
always satisfies $\lambda\lesssim\delta_{H}$, whereas the opposite
limit is never realized.

Substitution of Eq.\ (\ref{skyrmion-components-modified}) into Eq.\ (\ref{Lagrangian})
gives 
\begin{equation}
{\cal L}=\hbar\dot{\gamma}{\cal M}(\lambda)-E(\lambda,\gamma),
\end{equation}
where 
\begin{equation}
{\cal M}(\lambda)=\int dxdy\frac{2\lambda^{2}}{f^{2}(r)+\lambda^{2}}
\end{equation}
is the magnetic moment of the skyrmion and $E(\lambda,\gamma)$ is
the energy (\ref{E-continuous}) corresponding the modified profile
of the skyrmion.

The coupled equations of motion for ${\cal M}$ and $\gamma$ that
follow from Eq.\ (\ref{EL}) are 
\begin{equation}
\hbar\frac{d\gamma}{dt}\frac{d{\cal M}}{d\lambda}=\frac{dE}{d\lambda},\quad\hbar\frac{d\lambda}{dt}\frac{d{\cal M}}{d\lambda}=-\frac{dE}{d\gamma}.
\end{equation}
For small oscillations, $\delta\lambda$ and $\delta\gamma$, near
their equilibrium values $\delta\lambda_{1}$ and $\delta\gamma_{1}$,
approximating the energy by a parabolla, 
\begin{equation}
E(\lambda,\gamma)=E(\lambda_{1},\gamma_{1})+\frac{1}{2}E_{\lambda\lambda}\delta\lambda^{2}+\frac{1}{2}E_{\gamma\gamma}\delta\gamma^{2},
\end{equation}
one obtains linear equations of motion 
\begin{equation}
\frac{d\delta\gamma}{dt}=\frac{E_{\lambda\lambda}}{\hbar{\cal M}_{\lambda}}\delta\lambda,\quad\frac{d\delta\lambda}{dt}=-\frac{E_{\gamma\gamma}}{\hbar{\cal M}_{\lambda}}\delta\gamma,
\end{equation}
where $E_{\lambda\lambda}\equiv\partial^{2}E/\partial\lambda^{2}$,
$E_{\gamma\gamma}\equiv\partial^{2}E/\partial\gamma^{2}$, and ${\cal M}_{\lambda}\equiv d{\cal M}/d\lambda$.
These equations describe the oscillating motion of the dynamically
conjugate pair $\left\{ \delta\lambda,\delta\gamma\right\} $ at a
frequency 
\begin{equation}
\omega=\sqrt{\frac{E_{\lambda\lambda}E_{\gamma\gamma}}{\hbar{\cal M}_{\lambda}}}.\label{semi-analytical}
\end{equation}
This solution is more general than the one given above and it reproduces
Eq. (\ref{omega1lambda1}) if the BP skyrmion profile is used.

\begin{figure}
\centering{}\includegraphics[width=80mm]{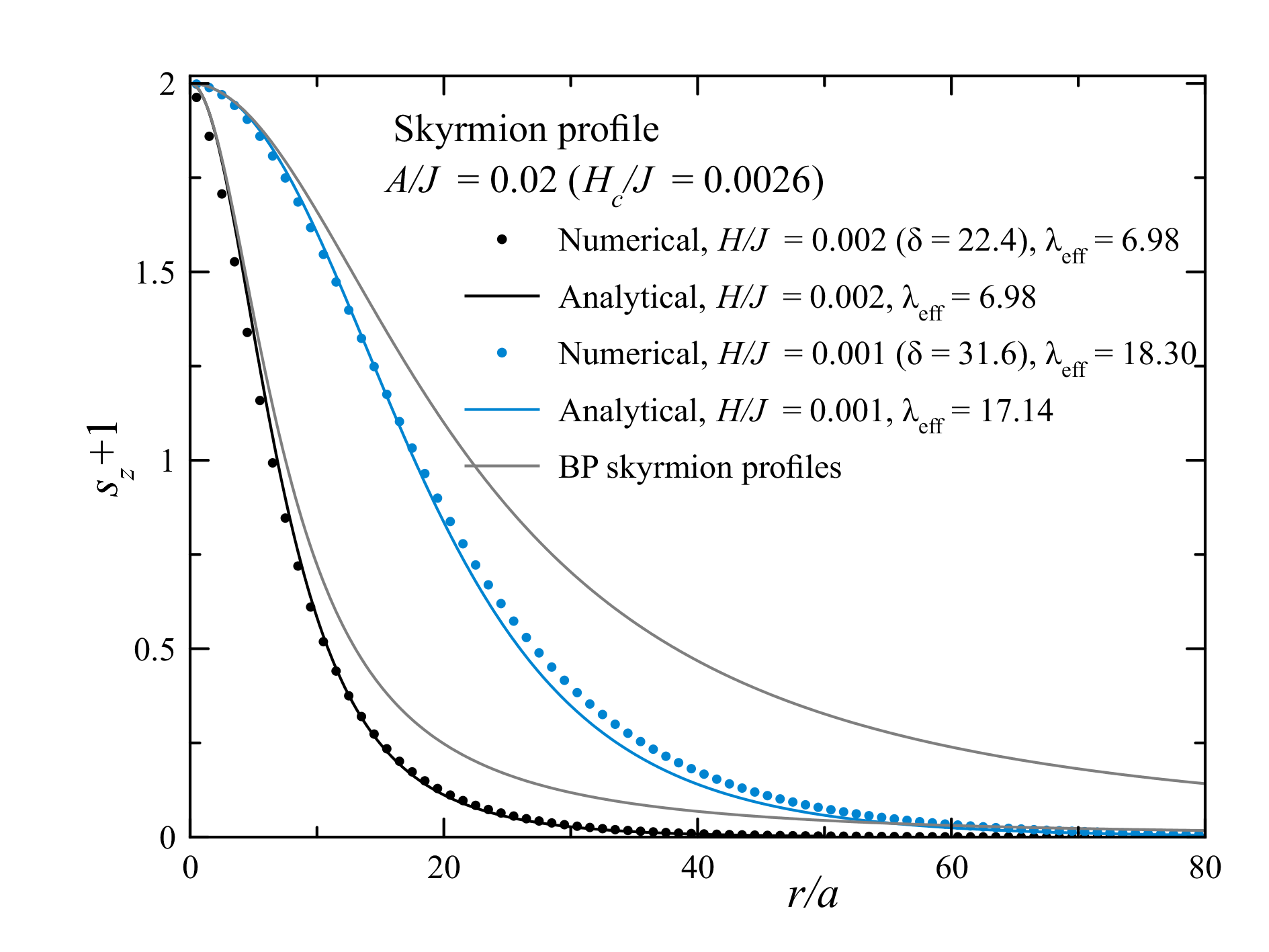}\caption{Modified skyrmion shape given by Eq.\ (\ref{skyrmion-components-modified})
with the equilibrium size $\lambda_{{\rm eff}}$ obtained by the numerical
minimization of the energy for two values of the magnetic field. Comparison
with the BP shape and numerical results obtained on the lattice are
also shown.}
\label{Fig-semi-analytical_profiles} 
\end{figure}

The energy $E(\lambda,\gamma)$ should now be computed numerically
with the help of Eqs. (\ref{E-continuous}), and (\ref{skyrmion-components-modified})
and minimized with respect to $\lambda$ and $\gamma$ to obtain their
equilibrium values. This gives $\gamma=\pi/2$ as before. The corresponding
shape and equilibrium size of the skyrmion computed that way and compared
with numerical results on the lattice are illustrated in Fig.\ \ref{Fig-semi-analytical_profiles}.
Deviation from the BP shape at large distances from the center of
the skyrmion is quite significant while disagreement between our semi-analytical
model and numerical results obtained on the lattice is rather small. 

A better agreement with numerical results on the lattice (achieved
in the absence of any fitting parameter) can also be seen in the plot
of $\omega(H)$ of Eq.\ (\ref{semi-analytical}) shown by red line
in Fig. \ref{Fig-omega-H+theory}. In particular, the semi-analytical
approach provides a much better description of the collapse region
than the crude analytical approach. However, the correct slope 0.8
in the low-field $\omega(H)$ is not captured. 

\section{Large-amplitude breathing mode\label{Sec-Large-amplitude-breathing-mode}}

\begin{figure}
\begin{centering}
\includegraphics[width=8.8cm]{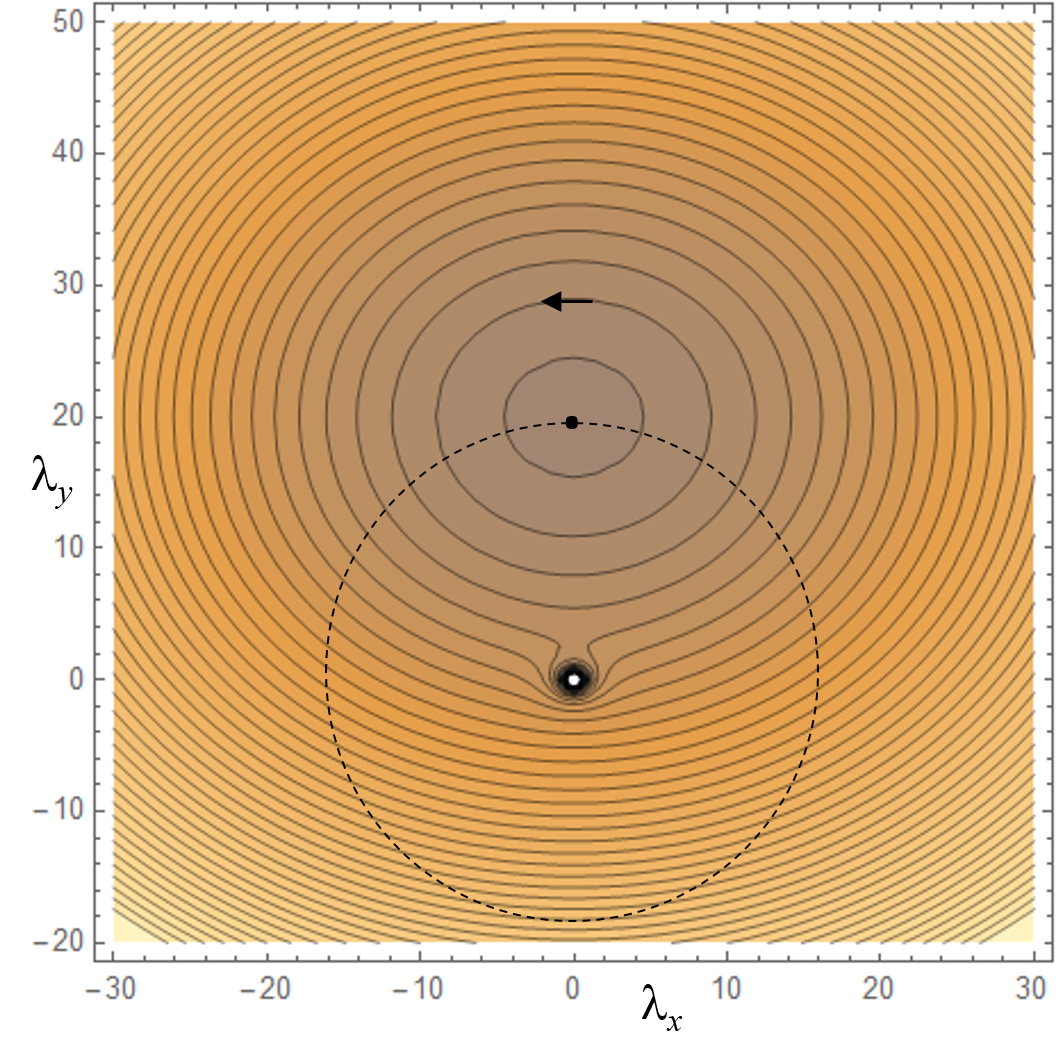}
\par\end{centering}
\caption{Equipotential lines in the $\left\{ \lambda_{x},\lambda_{y}\right\} =\left\{ \lambda\cos\gamma,\lambda\sin\gamma\right\} $
plane corresponding to Eq.\ (\ref{E-l}) with $l={\rm const}$.}

\label{Fig-ELandscape_lambda_x-lambda_y}
\end{figure}
\begin{figure}
\begin{centering}
\includegraphics[width=8.8cm]{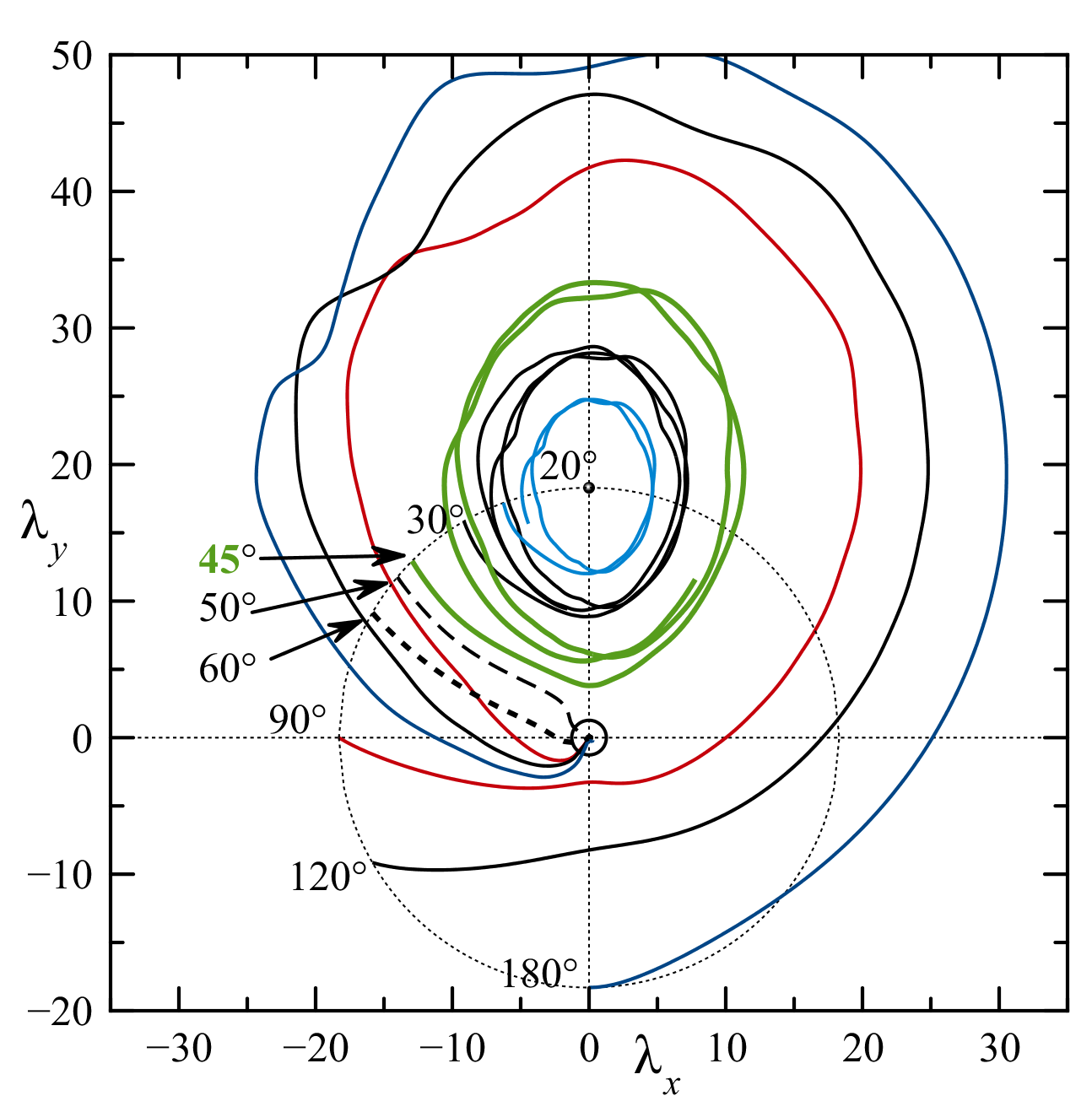}
\par\end{centering}
\caption{Dynamics of the breathing mode after rotation of the spins by a large
angle $\Delta\gamma$ in the system of $500\times500$ spins with
$A/J=0.02$ and $H/J=-0.001$. For $\Delta\gamma\apprge50\protect\textdegree$,
the skyrmion collapses. }

\label{Fig-lambda_x_lambda_y_motion}
\end{figure}

In this section, we investigate numerically the dynamics of large
amplitude breathing mode inititiated by rotating the spins by a large
angle $\Delta\gamma$. Instead of the phase diagram in terms of $\left\{ \lambda.\gamma\right\} $
shown in Fig. \ref{Fig-equipotential}, it is more convenient to use
the phase diagram in terms of $\left\{ \lambda_{x},\lambda_{y}\right\} =\left\{ \lambda\cos\gamma,\lambda\sin\gamma\right\} $.
For the same parameters as Fig. \ref{Fig-equipotential}, this new
phase diagram is shown in Fig. \ref{Fig-ELandscape_lambda_x-lambda_y}.
Here, all equipotential lines are closed. Oscillating regime corresponds
to equipotential lines that do not enclose the center $\left\{ 0,0\right\} $
where the skyrmion collapses. Rotating regime corresponds to the lines
that do enclose the center. Rotation of the spins by the angle $\Delta\gamma$
out of the equilibrium point denoted by the dot, corresponds to the
displacement along the dashed circle around the center. 

Numerical results in Fig. \ref{Fig-lambda_x_lambda_y_motion} show
that stable breathing motion is possible for $\Delta\gamma<45\textdegree$,
although this motion is clearly affected by the coupling to other
modes. For larger amplitudes, the skyrmion collapses approaching the
collapse point either directly ($\Delta\gamma=50\textdegree$ and
$60\textdegree$) or after one rotation ($\Delta\gamma=90\textdegree$,
$120\textdegree$, and $180\textdegree$). Such a quick dissipation
of the energy of the breathing mode should be due to the energy transfer
into the other modes, whereas the total energy of the system is conserved.
In fact, as the measured frequency of the large-amplitude breathing
modes approaches the FMR frequency ($\omega=0.000929$, 0.000988,
and 0.0010649 for $\Delta\gamma=90\textdegree$, $120\textdegree$,
and $170\textdegree$, respectively), energy transfer into bulk precession
mode becomes possible. This is not surprising because the rotational
breathing mode evolves in the same direction, thus it can efficiently
drive the bulk precession mode, losing its energy. To the contrary,
motion of $\gamma$ back and forth in the oscillating regime cannot
drive the bulk precession mode. When the skyrmion collapses, its entire
energy is converted into that of spin waves. 

\section{Conclusions}

\label{Sec-conclusions}

We have studied the breathing mode of a skyrmion stabilized by the
Dzyaloshinskii-Moriya interaction in a non-centrosymmetric magnetic
film. Compared to the previous studies our focus has been on large
fields close to the stability threshold. In that region the frequency
of the breathing mode reaches maximum on the field and then tends
to zero on approaching the collapse field. This effect must exist
not only for individual skyrmions but also for skyrmions forming a
lattice. It should not be difficult to test in experiment.

The above behavior of the breathing mode has been obtained by three
methods: Computations on lattices up to $500\times500$ in size; crude
analytical approximation using the Belavin-Polyakov shape of the skyrmion;
and a semi-analytical dynamical model based upon modified skyrmion
shape. All three models have produced qualitatively same behavior
and agree with each other quantitatively within $20\%$. Analytical
approach provides a simple picture of the breathing mode as coupled
oscillations of the skyrmion size and magnetic moment.

The most accurate numerical model on the lattice shows that the frequency
of the breathing mode in the oscillating regime is always below the
excitation spectrum in the bulk and, thus, it cannot decay into spin
waves. This explains why no damping of the breathing oscillation of
the skyrmion has been observed in the numerical experiment within
the conservative spin model. In real experiments the damping may result
from coupling to phonons and conducting electrons. It is expected
to be weak in insulating materials. On the other hand, in the rotating
regime the frequency of the breathing mode increases and it becomes
strongly damped apparently via energy transfer into the bulk precession
mode. Unlike the long-wavelength undamped micromagnetic approach,
the lattice-based nonlinear dynamics used here captures all scales
of excitations up to the atomic scale ($ka\sim1$), thus it describes
the processes of natural damping in the absence of damping added \textquotedblleft by
hand\textquotedblright . In that approach the short-wavelength spin
waves serve as the main energy reservoir for the relaxation. 

Current induced spin torques have been used to manipulate skyrmions
\cite{Fert-Nature2017}. In such experiments the breathing dynamics
could be used to control skyrmion trassport through a constriction
\cite{Hoffmann-PhysRep2017}. If skyrmions are utilized for data storage
and processing, our theory developed for the smallest skyrmions must
be useful for analyzing their response to external perturbations.

\section{Acknowledgments}

This work has been supported by the grant No. DE-FG02-93ER45487 funded
by the U.S. Department of Energy, Office of Science.

\end{document}